# Observation of Antiferroelectric Domain Walls in a Uniaxial Hyperferroelectric

Michele Conroy,[1,#,*] Didrik René Småbråten,[2,3,†,*] Colin Ophus,[4,*] Konstantin Shapovalov,[5] Quentin M. Ramasse,[6,7] Kasper Aas Hunnestad,[2] Sverre M. Selbach,[2] Ulrich Aschauer,[3,8] Kalani Moore,[9] J. Marty Gregg,[10] Ursel Bangert,[11] Massimiliano Stengel,[5,12] Alexei Gruverman,[13#] Dennis Meier[2,#]

[#]Corresponding author email: mconroy@imperial.ac.uk, agruverman2@unl.edu, dennis.meier@ntnu.no

[1] Department of Materials, London Centre of Nanotechnology, Imperial Henry Royce Institute, Imperial College London, SW7 2AZ, UK
[2] Department of Materials Science and Engineering, NTNU Norwegian University of Science and Technology, NO-7491 Trondheim, Norway
[3] Department of Chemistry, Biochemistry and Pharmaceutical Sciences, University of Bern, Switzerland
[4] National Center for Electron Microscopy, Molecular Foundry, Lawrence Berkeley National Laboratory, USA
[5] Institut de Ciencia de Materials de Barcelona (ICMAB-CSIC), Campus UAB, 08193 Bellaterra, Spain
[6] School of Physics and Astronomy, School of Chemical and Process Engineering, University of Leeds, UK
[7] SuperSTEM, SciTech Daresbury Science and Innovation Campus, Daresbury UK
[8] Department of Chemistry and Physics of Materials, University of Salzburg, 5020 Salzburg, Austria
[9] Direct Electron LP, San Diego, CA 92128, USA
[10] Centre for Quantum Materials and Technologies, School of Mathematics and Physics, Queen's University Belfast, Belfast, BT7 1NN, UK
[11] Department of Physics, Bernal Institute, University of Limerick, Ireland
[12] Institució Catalana de Recerca i Estudis Avançats (ICREA), Pg. Lluís Companys, 23 08010 Barcelona, Spain
[13] Department of Physics and Astronomy, Nebraska Center for Materials and Nanoscience, University of Nebraska, Lincoln, USA

[*] These authors contributed equally to this work

[†] Present address: Sustainable Energy Technology, SINTEF Industry, Forskningsveien 1, NO-0373 Oslo, Norway

**Ferroelectric domain walls are a rich source of emergent electronic properties and unusual polar order. Recent studies showed that the configuration of ferroelectric walls can go well beyond the conventional Ising-type structure. Néel-, Bloch-, and vortex-like polar patterns have been observed, displaying strong similarities with the spin textures at magnetic domain walls. Here, we report the discovery of antiferroelectric domain walls in the uniaxial ferroelectric $Pb_5Ge_3O_{11}$. We resolve highly mobile domain walls with an alternating displacement of Pb atoms, resulting in a cyclic 180º flip of dipole direction within the wall. Density functional theory calculations reveal that $Pb_5Ge_3O_{11}$ is hyperferroelectric, allowing the system to overcome the depolarization fields that usually suppress antiparallel ordering of dipoles along the longitudinal direction. Interestingly, the antiferroelectric walls observed under the electron beam are energetically more costly than basic head-to-head or tail-to-tail walls. The results suggest a new type of excited domain-wall state, expanding previous studies on ferroelectric domain walls into the realm of antiferroic phenomena.**

Lead germanate, $Pb_5Ge_3O_{11}$, is a uniaxial ferroelectric material with a spontaneous polarization $P_S$ = 4.8 µC cm$^{-2}$ along the crystallographic [001] direction and an enantiomorphic structure (space group $P3$).[1,2] One of the most peculiar consequences of the enantiomorphism in $Pb_5Ge_3O_{11}$ is its polarization-dependent optical activity, which makes ferroelectric 180° domains discernible in polarized light.[2,3] Because of the unipolar structure of $Pb_5Ge_3O_{11}$, only 180° domains exist in this material, forming irregular patterns elongated along the polar [001] axis as shown in Figure 1a. An interesting feature of the associated domain walls is that they can be strongly inclined with respect to the polar axis, giving rise to head-to-head and tail-to-tail sections with unusual electronic and dynamical responses. For example, the nominally charged domain walls have been reported to form characteristic saddle points that remove the need for screening charges[4] and scanning probe experiments showed that they expand in length under application of an electric field.[5] These physical properties are incompatible with the formation of classical Ising-type walls, pointing towards a more complex domain wall structure. In this study, we use density functional theory calculations (DFT) to understand the microscopic origin of the ferroelectric polarization in $Pb_5Ge_3O_{11}$ and apply different scanning transmission electron microscopy (STEM) methods to resolve the atomic-scale structure of the nominally charged domain walls.

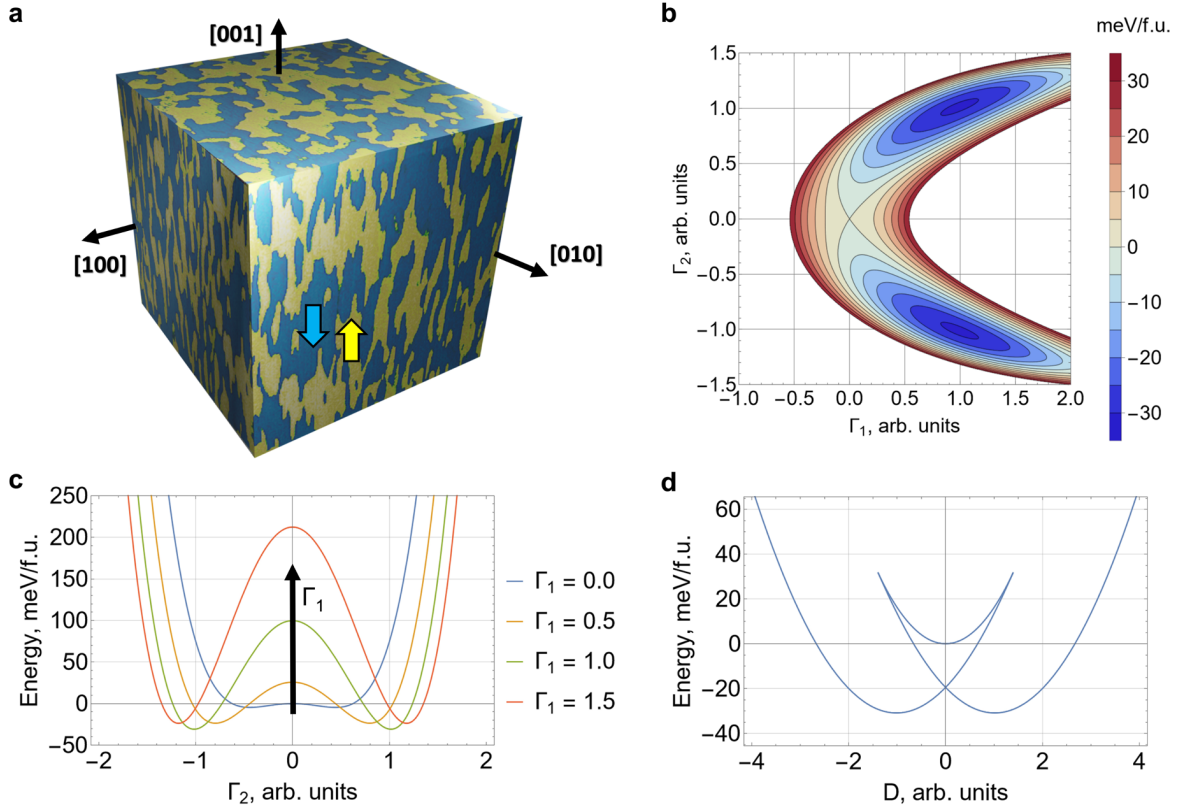

**Figure 1 | Domain structure and free energy landscape of Pb$_5$Ge$_3$O$_{11}$. a,** 3D sketch of the as-grown domain structure constructed based on piezoresponse force microscopy measurements. **b,** Contour plot of the two-dimensional potential energy landscape with respect to chiral $\Gamma_1$ and polar $\Gamma_2$ mode amplitudes as extracted from ab-initio calculations. **c,** Ferroelectric double-well potential for different chiral mode amplitudes $\Gamma_1$. **d,** Plot of the potential energy vs. displacement field, $D$, showing the hyperferroelectric behavior of the material with three possible polar states. The minima at $D = \pm 1$ correspond to the two polar states $+P$ and $-P$ ($E = -30$ meV/f.u.), which remain stable in open-circuit conditions ($D = 0$, $E = -20$ meV/f.u.). The local minimum at $D = 0$ corresponds to the non-polar prototype phase ($E = 0$ meV/f.u.).

We begin by analyzing the structural instabilities of the prototypical centrosymmetric bulk phase (space group $P\bar{6}$) that drive the ferroelectric order in Pb$_5$Ge$_3$O$_{11}$. Our DFT calculations reveal a single discontinuous unstable phonon band at the centre of the Brillouin zone (Supplementary Figure S1), consistent with the system's uniaxial ferroelectricity and the associated reduction in space group symmetry ($P\bar{6} \rightarrow P3$). Interestingly, by comparing the ferroelectric and prototypical structures (phonon analysis performed with AMPLIMODES[6]), we find that both a polar ($\Gamma_2$) and a chiral mode ($\Gamma_1$) are involved in the ferroelectric phase transition (Supplementary Movie 1 and Figure S2 present a visualization of the $\Gamma_1$ and $\Gamma_2$ modes). The corresponding two-dimensional potential energy landscape is given in Figure 1b with cell parameters fixed to those of the ferroelectric structure. These results show that the ferroelectric ground state is not stabilized by condensing the polar $\Gamma_2$ mode, but becomes stable only when both the polar and chiral modes condense, analogous to an improper ferroelectric phase transition.[7,8] We further find that the energy lowering by subsequent relaxation of the lattice parameters is significantly smaller than the phonon mode condensation, suggesting that contributions from macroscopic elasticity are minor. Following the symmetry of the prototypical phase, the coupling of the polar and chiral modes at leading order in the amplitude is of the type $\Gamma_2^2\Gamma_1$. Based on the ab-initio calculations, we construct a minimal Landau model of the potential energy landscape $F(\Gamma_1, \Gamma_2)$ as presented in Figure 1c, showing two minima as function of $\Gamma_2$ that correspond to the two ferroelectric domain states with $+P$ and $-P$ (see Supplementary Note 1 for details). To describe the effect of the depolarizing electric fields, we plot the potential energy, including the electrostatic energy ($F(\Gamma_1, \Gamma_2)+F_{el}$), as a function of the electric displacement field $D$, regarding $D$ as an external parameter (Figure 1d). The resulting curve is very different from the double-well potential of classical proper ferroelectrics. The main difference is the emergence of a bi-stable state in open-circuit conditions ($D = 0$) with a nonzero polarization, whereas the unstable prototypical structure lies at substantially higher energy. This behavior identifies $Pb_5Ge_3O_{11}$ as a uniaxial hyperferroelectric material,[9] where the polar instability survives under longitudinal (in addition to transverse) electrical boundary conditions (Supplementary Figure S1). Thus, in contrast to classical proper ferroelectrics, where depolarizing fields completely suppress the ferroelectric instability in open-circuit conditions, $Pb_5Ge_3O_{11}$ sustains its ferroelectric order for $D = 0$. Most importantly for this work, in principle, the hyperferroelectricity enables atomically sharp charged domain walls (head-to-head and tail-to-tail) even in the absence of mobile screening carriers as previously discussed for $LiNbO_3$,[10,11] making it largely immune against the emergence of uncompensated bound charges.

To measure the domain wall structure in $Pb_5Ge_3O_{11}$, we image the polar order at the atomic scale using different STEM experiments. The STEM images in Figures 2a-c delineate the domain configuration on the non-polar (010) surface, where the polarization direction within the field of view was determined by STEM differential phase contrast (DPC) (see Methods and Supplementary Figures S3 and S4).[12-14] Here, dark and bright areas correspond to $+P$ and $-P$ domains, respectively, revealing an extended tail-to-tail domain wall with a sawtooth-like structure. The image series shows that the tail-to-tail domain wall is remarkably mobile, changing position from frame to frame, which we attribute to an interaction with the electron probe.[13,15,16] The same behavior is observed in large-area scans obtained by scanning electron microscopy (SEM), independent of the scan direction of the SEM probe (Supplementary Figure S4).

Corresponding high-resolution annular dark-field (ADF) STEM images recorded along the [010] and [100] zone axes are shown in Figures 2d and 2e, respectively, presenting the atomic structure, which is consistent with earlier neutron diffraction studies.[17] $Pb_5Ge_3O_{11}$ contains layers of isolated tetrahedra ($GeO_4$) and layers of double tetrahedra ($Ge_2O_7$); the Pb atoms are located in threefold triangular pyramids and sixfold triangular prisms. The spontaneous polarization is associated with a displacement of Pb atoms located in triangular prisms along the [001] direction, resulting in distinct Pb plane spacings ($d_1$, $d_2$, and $d_3$) as illustrated in the unit cell images in Figures 2d,e. In $+P$ ($-P$) domains, three of the sixfold coordinated Pb atoms in the unit cell move upwards (downwards), leading to a shorter (longer) distance $d_1$ compared to $d_2$, whereas $d_3$ remains constant (see also Supplementary Figure S5). Quantifying the Pb plane spacings for a $-P$ domain, we find $d_1 = 3.462 \pm 0.021$ Å, $d_2 = 3.417 \pm 0.025$ Å, and $d_3 = 3.771 \pm 0.025$ Å. The relative change in Pb column positions, as well as the positions of light oxygen atoms relative to heavier Pb and Ge atomic columns (see Supplementary Figure S6 and S7), are in qualitative agreement with the reported ferroelectric structure of $Pb_5Ge_3O_{11}$[17] and our DFT results (Supplementary Figure S8). We note that quantitatively, however, the difference between $d_1$ and $d_2$ measured based on the ADF STEM is much smaller compared to the calculated values and previous experimental results.

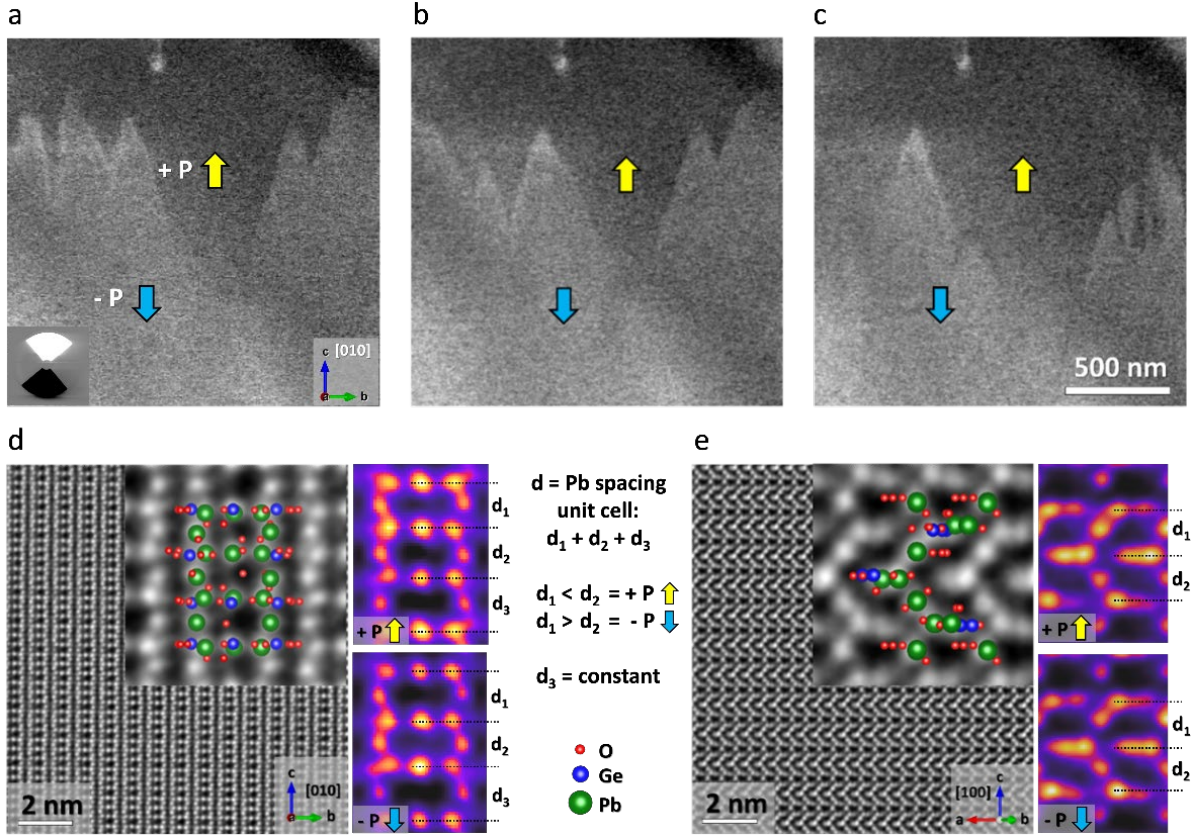

**Figure 2 | Polar structure of Pb$_5$Ge$_3$O$_{11}$. a-c,** STEM DPC maps recorded in consecutive frames, showing antiparallel 180º domains (+*P* dark, -*P* bright). The position of the domain walls varies between the three images, showing that the walls are highly mobile, moving under electron beam irradiation. The inset image of the quadrant ADF detector to **a** shows that the upper quadrant was subtracted from the lower to create the images **a-c**. Yellow and blue arrows indicate the polarization directions in +*P* and -*P* domains. **d,** Representative ADF STEM image of the atomic structure viewed along the [010] direction. The inset shows the atomic model of the unit cell, overlaid on a magnified section of the image[17]. The colorised ADF STEM panels on the right in **d** display the Pb spacing of unit cells associated with positive (+*P*, top) and negative (-*P*, bottom) polarization, parametrized by $d_1$, $d_2$, and $d_3$. **e,** Same as in **d** recorded along the [100] zone axis.

After characterization of the +*P* and -*P* domains, we next turn to the nominally charged domain walls. We find that bright field TEM imaging at low dose and low magnification reduces effects from domain wall motion, showing a region of reduced intensity between +*P* and -*P* domains with a width ≳ 16 nm (Supplementary Figure S9). This value is substantially larger than what is expected for a domain wall in a hyperferroelectric system[10,11] and also much larger than for 180º domain walls in conventional perovskite-type ferroelectrics, which have a typical width of about 5–10 Å.[18,19]. The large width prohibits imaging of the entire region of interest in a single STEM frame with spatial resolution and signal-to-noise ratio per atomic column sufficient for atomic displacement mapping. To overcome this experimental challenge, we perform high-resolution ADF STEM imaging separately on both sides of the low-intensity region associated with the domain wall as shown in Figures 3a,b (see Methods for details). The ADF STEM imaging demonstrates that a distinct stripe-like pattern arises in connection with the tail-to-tail domain wall, which distinguishes it from the +$P$ (Figure 3a) and -$P$ (Figure 3b) domains with homogeneous polarization orientation. This stripe-like pattern is observed for different polarization configurations as shown in Supplementary Figure S10, suggesting that it arises in association with both tail-to-tail and head-to-head domain walls. The stripe-like pattern is more readily illustrated using Fourier filtering, as shown in the processed Fourier-filtered image in Figure 3c (see Supplementary Figure S11 and Ref.[20] for details).

We then map the atomic displacements as explained in the Methods section and fingerprint the local polarization as shown in Figures 3d-f. We find that the atomic structure of the domain walls typically exhibits three characteristic regions. Starting from the -$P$ domain in Figure 3b with uniform orientation of the electric dipoles (Figure 3d), the atomic structure transforms into an aperiodic pattern of dipole alignment formed by unit cells with limited short-range ordering of the alternating Pb displacements (Figure 3e). Subsequently, this aperiodic pattern evolves into a well-defined structure with periodically changing directions of Pb displacements, forming a regular antiparallel dipole alignment (Figure 3f). We note that this type of structural transformation is independent of the viewing direction and wall type (i.e., it is also seen at head-to-head domain walls) and can be observed in the images recorded along the [010] and [100] directions (Supplementary Figure S10). Based on the STEM data, we conclude that toward the center of nominally charged tail-to-tail and head-to-head domain walls, the ferroelectric order transforms into an antipolar structure, corresponding to antiferroelectrically coupled dipole layers (Figure 3g).

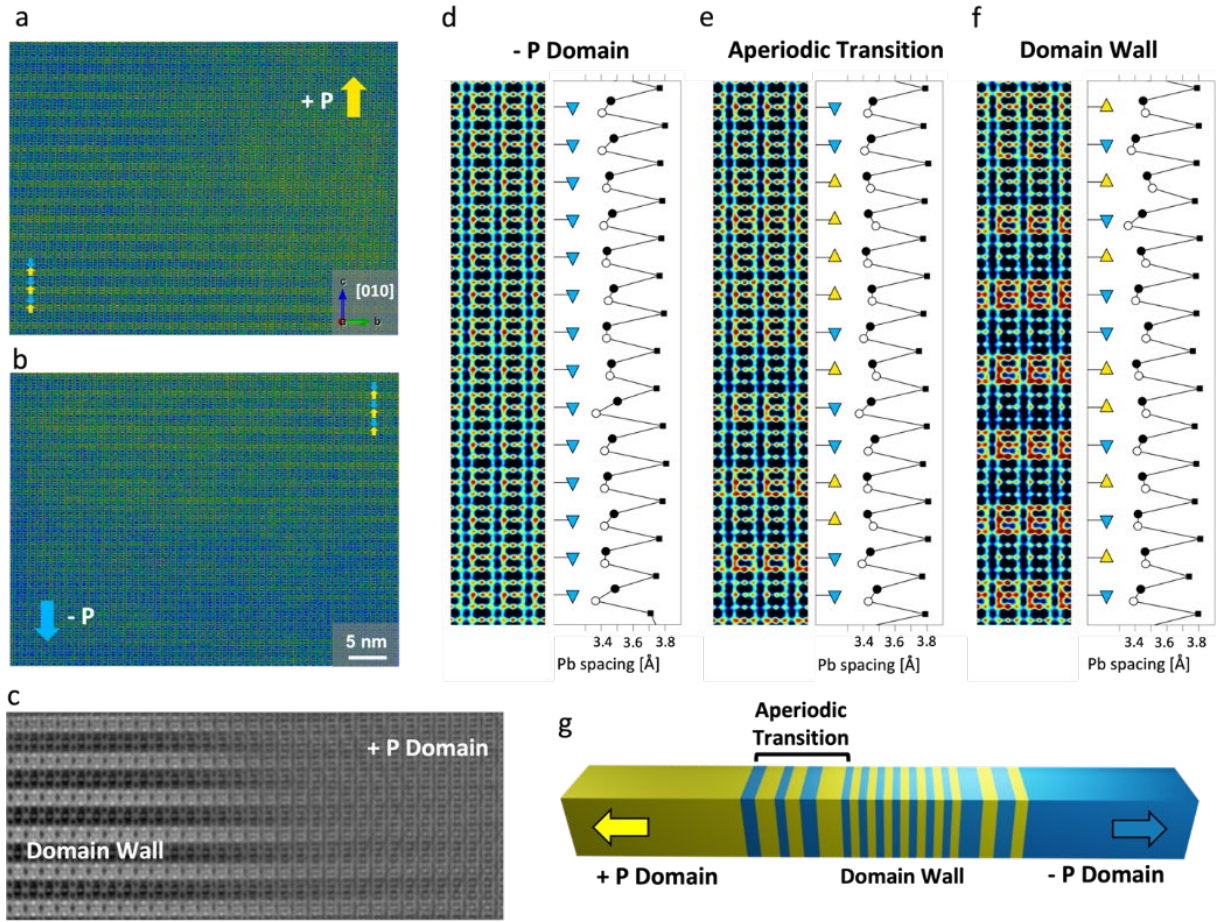

**Figure 3 | Atomic-scale structure of domain walls in $Pb_5Ge_3O_{11}$. a,b,** ADF STEM imaging of +*P* (**a**) and -*P* (**b**) domain regions transitioning into anti-polar regions; imaged along the [010] direction. **c,** Fourier-filtered image (selecting the superlattice spots) of a section of the ADF STEM image in **a**, highlighting the stripe-like pattern associated with the tail-to-tail domain wall. **d-f,** ADF STEM maps along with the corresponding cell-to-cell dipole alignments arising due to the displacement of the Pb atoms. **d,** +*P* domain region, **e,** aperiodic transition region, **f,** domain wall. Black and white circles correspond to the measured atomic spacing of the Pb atoms in the triangular prisms in the [001] direction. Black squares correspond to the unshifted Pb atoms in the triangular pyramids. Yellow and blue arrow heads denote the resulting +*P* and -*P* polarizations based on the inter-layer distances $d_i$ ($i$ = 1,2,3) as defined in Figure 2 (+*P*: $d_1 < d_2$; -*P*: $d_1 > d_2$). **g,** Schematic illustration of the dipole layer-by-layer alignment corresponding to the atomic structure in **d** to **f**.

The observation of antiparallel dipoles within the tail-to-tail and head-to-head walls is consistent with the robustness of $Pb_5Ge_3O_{11}$ against the emergence of domain wall bound charges as expected due to its hyperferroelectric nature. The hyperferroelectricity alone, however, cannot explain the short-period modulation observed experimentally. As structural relaxations of larger supercells with modulated dipolar order confirm (Supplementary Figure S12), a single and atomically sharp wall is the energetically favorable ground state, consistent with earlier theoretical works on hyperferroelectrics.[10] This leads us to the conclusion that hyperferroelectricity plays a key role concerning the building blocks of the observed antiferroelectric domain walls in $Pb_5Ge_3O_{11}$, as it enables the formation of atomically sharp antiparallel segments. The aperiodic and periodic combinations of such segments as seen in Figure 3, however, represent excited states that require an additional driving force. Likely candidates are temperature-driven effects and dynamical phenomena caused by the electron beam, in agreement with the reduced ferroelectric distortions measured by ADF STEM (Figures 2d,e) and the electron-beam induced domain wall motion seen in Figures 2a–c (and Supplementary Figures S3 and S4).

To gain additional insight into the electronic properties of the emergent antiferroelectric domain walls, we model their structure with an antiferroelectric supercell (Figure 4a) and calculate the corresponding potential profile (Figure 4b) and local density of states (DOS, Figure 4c) using DFT. The calculations reveal that the antiferroelectric structure leads to a flattening of the electrostatic potential and corresponding band-energy shifts in the DOS compared to individual head-to-head or tail-to-tail domain walls (Supplementary Figure S12). Importantly, the results in Figures 4b,c indicate that the conductivity at the antiferroelectric domain walls is similar to the bulk despite the bound charges that exist within the wall (i.e., similar band gap and no additional defect states). The latter is in agreement with previous local transport measurements, which demonstrated that the nominally charged domain walls in $Pb_5Ge_3O_{11}$ exhibit the same electronic conductance as the surrounding domains[5].

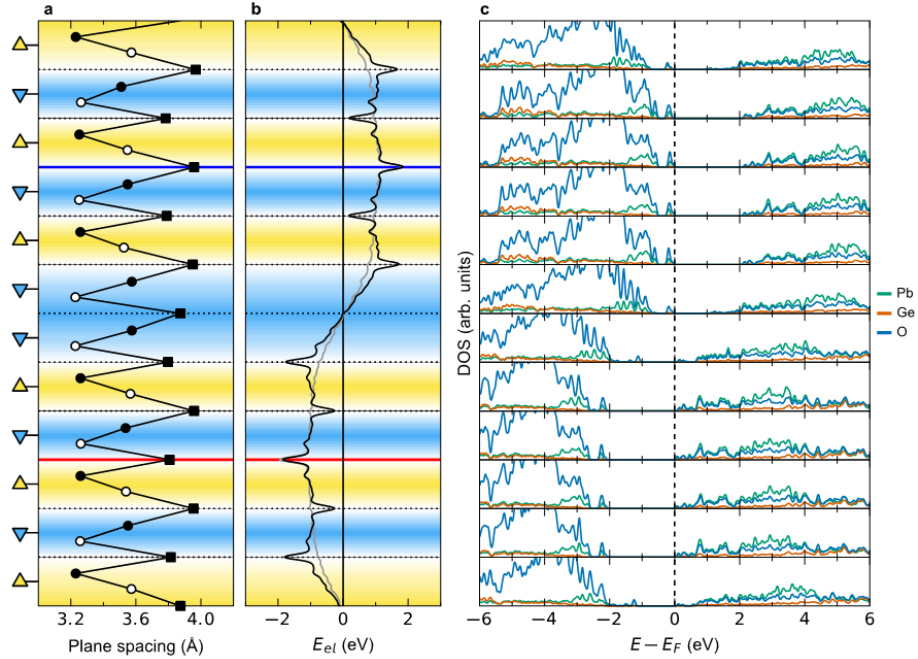

**Figure 4 | DFT modeling of the antiferroelectric domain wall structure in $Pb_5Ge_3O_{11}$. a-c,** Calculated Pb plane spacing (**a**), local potential profile (**b**), and electronic density of states (DOS) (**c**) across the supercell. Yellow and blue arrow heads in **a** correspond to $+P$ and $-P$ polarizations, respectively, whereas closed circles, open circles, and squares correspond to the plane spacings $d_1$, $d_2$, and $d_3$, respectively. For comparison, the potential for a conventional domain wall configuration (Supplementary Figure S12) is plotted in grey in **b**. The horizontal lines in **a** and **b** illustrate positions of interfaces between unit cells of opposite polarization direction, where the red and blue lines mark the center of the charged domain walls. The vertical dashed line in **c** corresponds to the Fermi level.

The observation of anti-polar order at ferroelectric domain walls extends previous work into the realm of antiferroic phenomena. Interestingly, the domain-wall-related evolution from ferroelectric to antiferroelectric order through an aperiodic structure is reminiscent of frustrated magnets. In the latter case, competing magnetic exchange interactions drive the system away from simple ferromagnetic arrangements, promoting incommensurate/aperiodic spin textures and antiferromagnetic order. In particular, the obtained results introduce an innovative strategy for designing antipolar structures, which can be applied also in heterostructures and superlattices: by utilizing hyperferroelectrics in combination with strain and electrostatic boundary conditions - analogous to the conditions found at the charged ferroelectric domain walls - competing polar phases may be exploited to create and fine-tune antiferroelectric 2D systems. This possibility may lead to artificial antiferroelectrics and ferroelectric-antiferroelectric superlattices with unusual physical properties. The engineering of artificial antiferroelectrics and (hyper)ferroelectrics with high-density antiferroelectric domain walls represents a promising approach for increasing the currently limited number of known antiferroelectric single-phase materials and antiferroelectric systems with improved functionality.

## Methods

**Sample Preparation.** Electron transparent cross-sections of single crystal $Pb_5Ge_3O_{11}$ were prepared for STEM using a dual-beam focused ion beam (FIB) integrated SEM (Thermo-Fisher Scientific FEI Helios G4 CX). The specimens were mounted onto Omniprobe copper-based lift-out grids. Thinning of the samples was done with decreasing accelerating voltage and electron beam current[22] in four steps: (1) thinning from 2 µm to 800 nm was performed using a 30 keV, 0.23 nA Ga ion beam, (2) 800 to 500 nm: 16 keV, 50 pA, (3) 500 to 300 nm: 8 keV, 50 pA, and (4) below 100 nm: 5 keV, 46 pA. Final polishing was done with 2 keV and 9 pA. Plan view FIB samples were made for STEM imaging of the polar top facet using the method detailed by Ref [23].

**Scanning Electron Microscopy.** The switching behavior of the domains over large fields of view (i.e., tens of µm) on the surface of the bulk crystal of $Pb_5Ge_3O_{11}$ was analyzed by scanning electron microscopy (SEM) as shown in Supplementary Figure S4a,b. The data was recorded using a Thermo-Fisher Scientific Apreo SEM with 5kV acceleration voltage and 0.8 nA electron beam current. See, e.g., Ref[21] for a review on domain and domain wall contrast in SEM. As shown in Figure S4a,b, the domain pattern on the non-polar surface changes in every scan, independent of the scan direction of the SEM probe.

**Scanning Transmission Electron Microscopy.** STEM imaging was performed using a Thermo-Fisher Scientific double aberration-corrected monochromated Titan Themis Z, operated at 300 kV accelerating voltage. The convergence angle for ADF STEM imaging was 24 mrad and the collection angle was 52–200 mrad using the HAADF detector; the measured fluorescent screen current was ≈ 30 pA. Additional imaging was carried out using a Nion UltraSTEM 100MC "Hermes" microscope at a lower accelerating voltage of 60 kV, with a convergence semi-angle of 33 mrad and typical beam current of 30pA. The HAADF detector angular range was 85-180 mrad. STEM DPC mapping was done using a segmented ADF detector and processed via Thermo-Fisher Velox software. To capture the internal atomic-scale structure, initially a domain wall was imaged using low magnification STEM DPC. After several scans from the same direction, a charged domain wall became stable. Then the beam was blanked, and the magnification was increased to the required level with enough pixels per atomic column for post processing. After this, a single STEM ADF frame was captured. The presence of the domain wall pattern was readily identified by respective extra spots in the Fourier transform of the STEM image, which arise due to the alternating direction of Pb atoms displacement at the unit cell level (Supplementary Figure S11). If a longer camera length (and thus decreased collection angle) was used for ADF imaging, the domain wall was seen as an alternating stripe of high and low contrast, associated with the changing in the atomic column spacing as seen in Figures 3a-c. Series of consecutive STEM images with relatively short pixel dwell times were drift-corrected by first aligning and averaging all images in a time series and then calculating a mean unit cell. Next, this unit cell was tiled to fill the entire image, which was then used as a reference image for measuring and correcting the distortions using the procedure given in Ref [24]. Since absolute distances were not preserved, only relative site displacements were measured. To analyze the $Pb_5Ge_3O_{11}$ structure, we averaged the images both in one dimension (1D) perpendicular to the domain wall and in two dimensions (2D), to produce mean unit cells as described by Danaie et al.[25]. This averaging was performed by first fitting the peak positions and then assigning each peak to a lattice coordinate, and finally computing the best fit lattice. In the 1D and 2D averaged images, we fit each Pb site to a 2D Gaussian distribution to estimate its position. For the displacement plots shown in Figure 2, we assumed the total unit cell dimensions were equal to the bulk value. Transmission electron microscopy (TEM) using parallel beam mode and the Gatan Oneview camera was used to suppress effects from domain wall motion during imaging. The response of the sample, however, was dose-rate dependent and the threshold value was found to be too low to achieve atomic resolution, so that STEM imaging was performed instead. Dose measurements were acquired by taking the reading from the incident beam hitting the fluorescent screen in units of $e^- \text{ Å}^{-2}$, which was converted into a dose rate based on the exposure time of acquisition. The Pb-Pb plane spacing (i.e., $d_1$, $d_2$, and $d_3$) was estimated from 14 unit cells of the -$P$ domain shown in Figure 3d, with the error bars representing the standard deviation of these measurements.

**Density Functional Theory.** Density functional theory calculations were performed using VASP.[26-28] The domain walls were modelled by 684 atom supercells comprised of 12 unit cells along the lattice vector $c$. Since the domain walls are experimentally found to span far beyond what our DFT models can capture, two unit cells in each domain were fixed to the relaxed bulk structure to ensure bulk-like properties in the center of each domain. The domain walls were modelled by four unit cells accordingly. Pb ($5d$, $6s$, $6p$), Ge ($3d$, $4s$, $4p$), and O ($2s$, $2p$) were treated as valence electrons, with a plane-wave cutoff energy of 550 eV. Brillouin zone integration was performed using a Γ-centered 2x2x2 grid for unit cell calculations, and a 2x2x1 grid for the charged domain wall supercells. Lattice positions were relaxed until the residual forces on all the atoms were below 0.05 eV/Å, with lattice parameters fixed to relaxed bulk values. Ferroelectric polarization was determined using the Berry phase method,[29-31] performed on unit cells extracted from the relaxed domain wall supercells. Frozen phonon calculations[32] were performed on 2x2x2 supercells (456 atoms) with the finite displacement method as implemented in the PHONOPY code,[33] using atomic displacements of 0.01 Å for calculating the force constants. Phonon analysis was carried out using the AMPLIMODES[6] and Sumo[34] packages.

## Acknowledgements

M.C. acknowledges funding from Science Foundation Ireland (SFI) Industry Fellowship (18/IF/6282), Royal Society Tata University Research Fellowship (URF\R1\201318), EPSRC NAME Programme Grant EP/V001914/1 and Royal Society Enhancement Award RF\ERE\210200EM1. The research by M.C., K.M., J.M.G., U.B., and A.G. was supported by the US-Ireland R&D Partnership Programme (grant no. USI 120), National Science Foundation (NSF) grant DMR-1709237 and Science Foundation Ireland (16/US/3344). D.R.S. and S.M.S. acknowledge the Research Council of Norway (FRINATEK project No. 275139/F20) for financial support. D.R.S. and U.A. acknowledge the Swiss National Science Foundation (Project No. 200021_178791) for financial support. Computational resources were provided by UNINETT Sigma2 (project no. NN9259K), and UBELIX (http://www.id.unibe.ch/hpc) - the HPC cluster at the University of Bern. C.O. acknowledges support from the DOE Early Career Research Program. Work at the Molecular Foundry was supported by the Office of Science, Office of Basic Energy Sciences, of the U.S. Department of Energy under Contract No. DE-AC02-05CH11231. SuperSTEM is the U.K. National Research Facility for Advanced Electron Microscopy, supported by the EPSRC through grant EP/W021080/1. M.S. and K.S. acknowledge support from Ministerio de Ciencia Y Innovación (MICINN-Spain) through Grant No. PID2019-108573GB-C22; from Severo Ochoa FUNFUTURE center of excellence (CEX2019-000917-S); from Generalitat de Catalunya (Grant No. 2021 SGR 01519); and from the European Research Council (Grant Agreement No. 724529). D.M. thanks NTNU for support through the Onsager Fellowship Program, the Outstanding Academic Fellow Program, and acknowledges funding from the European Research Council (ERC) under the European Union's Horizon 2020 Research and Innovation Program (Grant Agreement No. 863691). The Research Council of Norway (RCN) is acknowledged for the support to the Norwegian Micro- and Nano-Fabrication Facility, NorFab, project number 295864.

## Contributions

M.C. prepared the FIB samples. STEM measurements were performed by M.C., with advice from C.O. and Q.R.; D.R.S. carried out DFT calculations supervised by S.M.S. and U.A. M.C. and C.O. performed STEM data analysis. K.A.H. performed SEM measurements and contributed to the basic structure analysis, supervised by D.M. K.S. analysed the DFT results and the antiferroelectric ordering supervised by M.S. J.M.G. analysed the domain wall structure. D.M. and A.G. initiated and coordinated the project and wrote the manuscript together with M.C. and D.R.S. and with input from all authors. All authors discussed the results and contributed to the final version of the manuscript.

## Corresponding authors

Michele Conroy, mconroy@imperial.ac.uk

Alexei Gruverman, agruverman2@unl.edu

Dennis Meier, dennis.meier@ntnu.no

## Ethics declarations

Competing interests

The authors declare no competing interests.

# Supplementary Information

# Observation of Antiferroelectric Domain Walls in a Uniaxial Hyperferroelectric

Michele Conroy,[1,#,*] Didrik René Småbråten,[2,3,†,*] Colin Ophus,[4,*] Konstantin Shapovalov,[5] Quentin M. Ramasse,[6,7] Kasper Aas Hunnestad,[2] Sverre M. Selbach,[2] Ulrich Aschauer,[3,8] Kalani Moore,[9] J. Marty Gregg,[10] Ursel Bangert,[11] Massimiliano Stengel,[5,12] Alexei Gruverman,[13#] Dennis Meier[2,#]

[#]Corresponding author email: mconroy@imperial.ac.uk, agruverman2@unl.edu, dennis.meier@ntnu.no

[1] Department of Materials, London Centre of Nanotechnology, Imperial Henry Royce Institute, Imperial College London, SW7 2AZ, UK
[2] Department of Materials Science and Engineering, NTNU Norwegian University of Science and Technology, NO-7491 Trondheim, Norway
[3] Department of Chemistry, Biochemistry and Pharmaceutical Sciences, University of Bern, Switzerland
[4] National Center for Electron Microscopy, Molecular Foundry, Lawrence Berkeley National Laboratory, USA
[5] Institut de Ciencia de Materials de Barcelona (ICMAB-CSIC), Campus UAB, 08193 Bellaterra, Spain
[6] School of Physics and Astronomy, School of Chemical and Process Engineering, University of Leeds, UK
[7] SuperSTEM, SciTech Daresbury Science and Innovation Campus, Daresbury UK
[8] Department of Chemistry and Physics of Materials, University of Salzburg, 5020 Salzburg, Austria
[9] Direct Electron LP, San Diego, CA 92128, USA
[10] Centre for Quantum Materials and Technologies, School of Mathematics and Physics, Queen's University Belfast, Belfast, BT7 1NN, UK
[11] Department of Physics, Bernal Institute, University of Limerick, Ireland
[12] Institució Catalana de Recerca i Estudis Avançats (ICREA), Pg. Lluís Companys, 23 08010 Barcelona, Spain
[13] Department of Physics and Astronomy, Nebraska Center for Materials and Nanoscience, University of Nebraska, Lincoln, USA

[*] These authors contributed equally to this work

[†] Present address: Sustainable Energy Technology, SINTEF Industry, Forskningsveien 1, NO-0373 Oslo, Norway

# Supplementary Notes

**Supplementary Note 1**

Based on our ab-initio calculations, we construct a minimal Landau model of the potential energy landscape, which is given by

$$F(\Gamma_1, \Gamma_2) = \frac{1}{2}a_1 p^2 + \frac{1}{4}a_{11} p^4 + \frac{1}{2}b_1 q^2 + \frac{1}{4}b_{11} q^4 + c p^2 q + d p^2 q^2 \quad (1)$$

where $p$ and $q$ are the amplitudes of the polar mode ($\Gamma_2$) and the chiral mode ($\Gamma_1$), respectively, normalized so that the energy minimum is located at $q = 1, p = \pm 1$. The parameters extracted via least-square fitting of the first-principles data points are summarized in Table 1.

**Table 1 |** Calculated Landau parameters of the free energy landscape in Eq. (1) obtained from DFT calculations.

| Parameter | Value (meV/f.u.) |
|---|---|
| $a_1$ | -95.5256 |
| $a_{11}$ | 490.127 |
| $b_1$ | 208.144 |
| $b_{11}$ | -17.2818 |
| $c$ | -225.962 |
| $d$ | 20.6906 |

Simultaneously with the energy, we extract the electrical polarization as a function of ($\Gamma_1, \Gamma_2$), which displays a linear behavior as a function of $\Gamma_2$ and negligible dependence on the inversion-even $\Gamma_1$ mode amplitude,

$$P = P_S p . \quad (2)$$

This allows us to incorporate the electrostatic energy (which is expected to play a crucial role in the observed head-to-head and tail-to-tail dipolar structures) in a simple form as[35]

$$F_{el} = \frac{\varepsilon E^2}{2} = \frac{(D-P)^2}{2\varepsilon}, \quad (3)$$

where ε is the dielectric permittivity excluding the $\Gamma_2$ contribution (we use the value $\varepsilon = 25\varepsilon_0$ extracted from ab-initio, where $\varepsilon_0$ is the vacuum permittivity), $P$ is the polarization mediated by $\Gamma_2$, and $D$ is the electric displacement field, which we regard as an external parameter. At this point, we extract the electrical phase diagram of bulk $Pb_5Ge_3O_{11}$ by plotting the stationary points of $F+F_{el}$ as a function of $D$, Figure 1d. Remarkably, the resulting curve looks very different compared to the simple double-well potential of other proper ferroelectrics. The main surprise consists in finding a bi-stable state in open-circuit conditions ($D = 0$) with a nonzero polarization, while the third $D = 0$ solution (corresponding to the unstable prototypical structure) lies at significantly higher energy. Such a peculiar behavior unambiguously identifies $Pb_5Ge_3O_{11}$ as a uniaxial hyperferroelectric.

To see this, one can set $D = 0$ and calculate the second derivative of $F+F_{el}$ with respect to the polarization at $P = 0$. We obtain

$$A' = \frac{\partial^2 (F+F_{el})}{\partial P^2} = \frac{a_1}{P_S^2} + \frac{1}{\varepsilon}, \qquad (4)$$

where the term $1/\varepsilon$ embodies the depolarizing fields contribution to the harmonic Landau coefficient $a_1/P_S^2 = -2.0 \times 10^{10}$ m/F (here we used the formula unit volume $3.2 \times 10^{-28}$ m$^{-3}$ that we observed in ab-initio). We find that $A' = -1.6 \times 10^{10}$ m/F is still negative, consistent with the formal definition of hyperferroelectricity: the polar instability survives under longitudinal (in addition to transverse) electrical boundary conditions. This is further confirmed by the calculated phonon band structure of the prototypical phase in Supplementary Figure S1, where the discontinuous phonon band remains unstable when approaching the $\Gamma$ point from either the transversal or longitudinal direction.

# Supplementary Figures

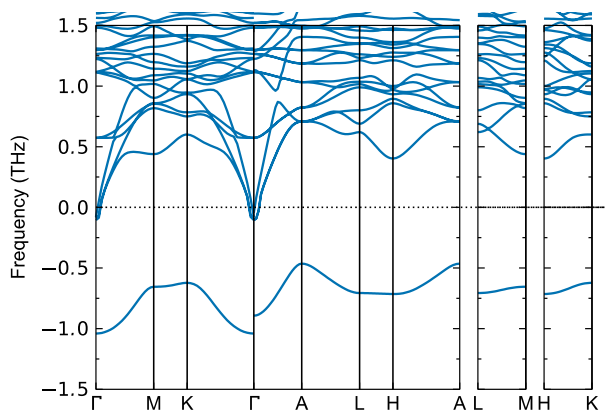

**Figure S1 | Phonon band structure of prototype Pb$_5$Ge$_3$O$_{11}$.** Calculated phonon band structure of prototype Pb$_5$Ge$_3$O$_{11}$, including non-analytical term correction. Hyperferroelectricity is confirmed by the instability of the discontinuous phonon band approaching the Γ point from either the transversal or longitudinal direction, as shown for the K-Γ-A segment.

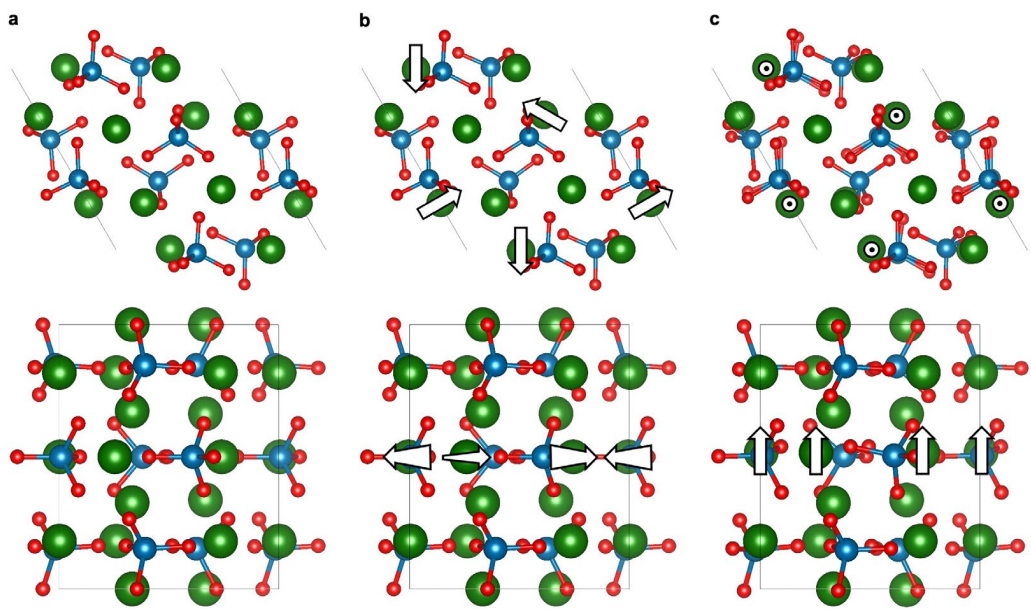

**Figure S2 | Visualization of phonon modes in prototype Pb$_5$Ge$_3$O$_{11}$. a-c,** Side and top views of prototype Pb$_5$Ge$_3$O$_{11}$ (**a**), the chiral Γ$_1$ phonon mode (**b**), and the polar Γ$_2$ phonon mode (**c**) associated with the ferroelectric phase transition. The white arrows in **b-c** illustrate the major displacements in the Pb6 plane for the two phonon modes relative to the prototype phase. The phonon modes are further visualized in Supplementary Movie 1.

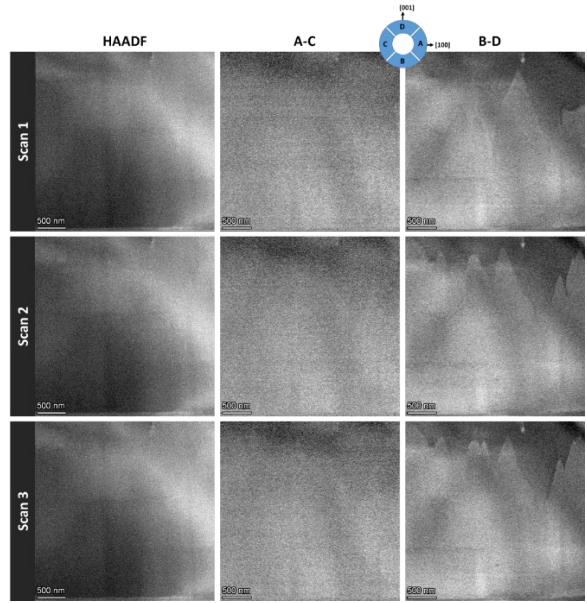

**Figure S3 | STEM DPC series.** Series of STEM scans with data collected on a HAADF detector, DF4 segments A and C, DF4 segments B and D. No clear deflection in the A-C direction and no clear change in contrast is observed in the HAADF data, consistent with the uniaxial polarization (B-D) of $Pb_5Ge_3O_{11}$.

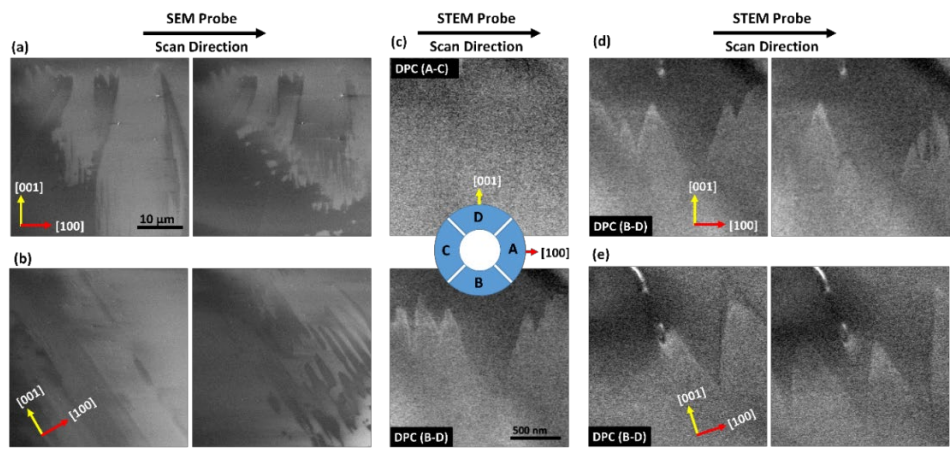

**Figure S4 | Domain motion in SEM and STEM DPC. a,b,** Series of SEM scans showing domain patterns which are elongated along specific crystallographic directions. **c,** STEM DPC mapping show no deflection in C-A direction, consistent with the the uniaxial nature of $Pb_5Ge_3O_{11}$. **d,e,** STEM DPC series, mapping the ferroelectric domanis with rotating scan direction.. A rotation of the incoming SEM or STEM probe does not change the direction/shape of the domains, which are typically elongated in the direction of the polar axis.

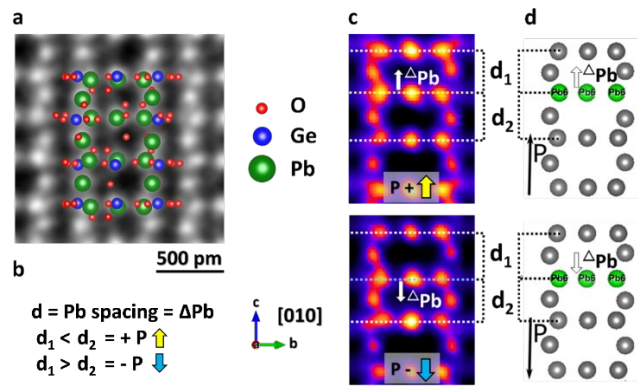

**Figure S5 | Atomic resolution STEM study of polarization vector. a,** Representative STEM ADF image of the atomic structure viewed along the [010] direction. The inset shows the atomic reconstruction of the unit cell. **b,** Relation between Pb plane spacing and polarization direction. **c**, STEM ADF showing how the Pb plane spacings $d_1$ and $d_2$ are determined. **d**, Atomic reconstruction corresponding to the data in **c**.

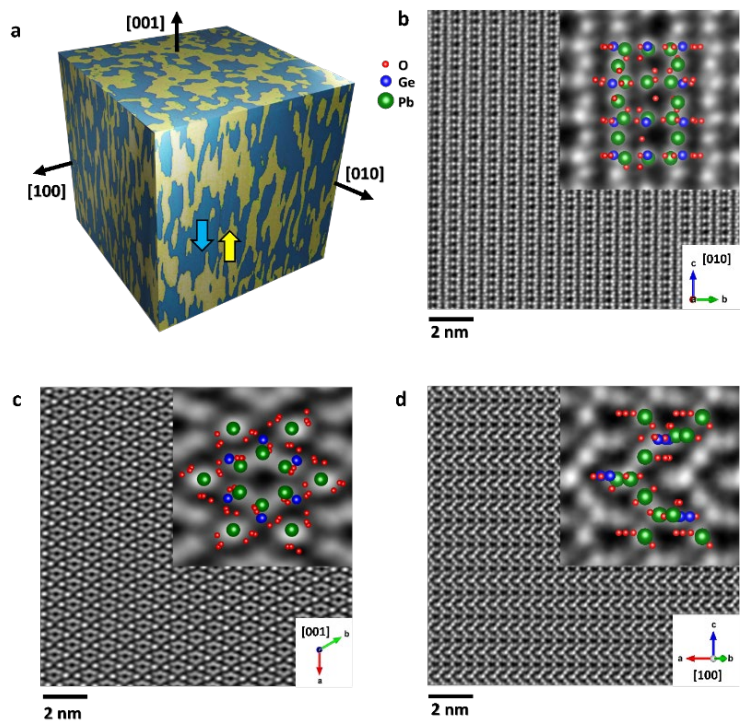

**Figure S6 | Atomic resolution analysis of crystallographic directions. a,** Composite sketch of the as-grown domain structure in $Pb_5Ge_3O_{11}$ revealed by PFM. **b-d,** Atomic resolution STEM ADF image recorded along the three major zones [001], [010] and [100] with atomic models.[17]

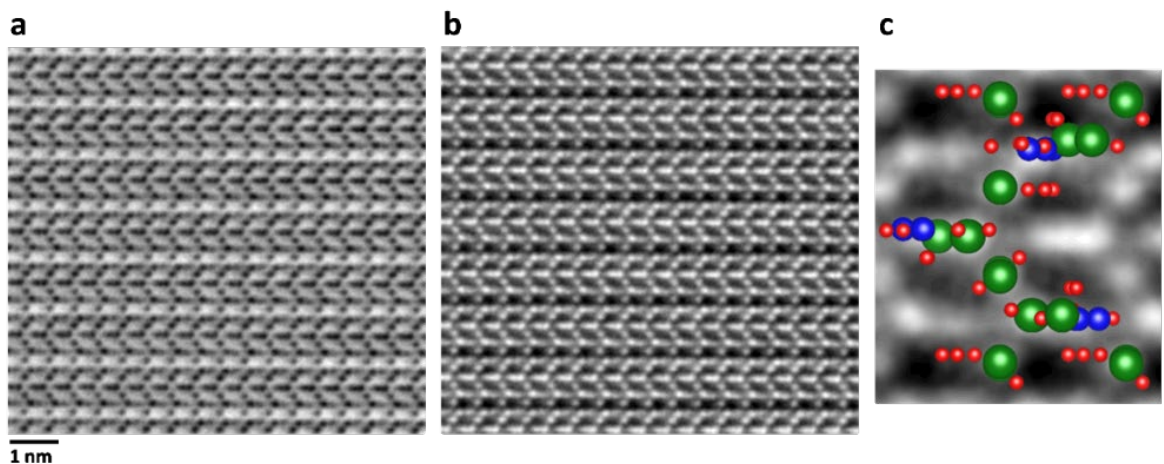

**Figure S7 | Atomic analysis of light and heavy elements. a,** STEM BF scan. **b,** Inverted STEM BF scan. **c,** Atomic unit cell model on top of experimental an inverted STEM BF scan.

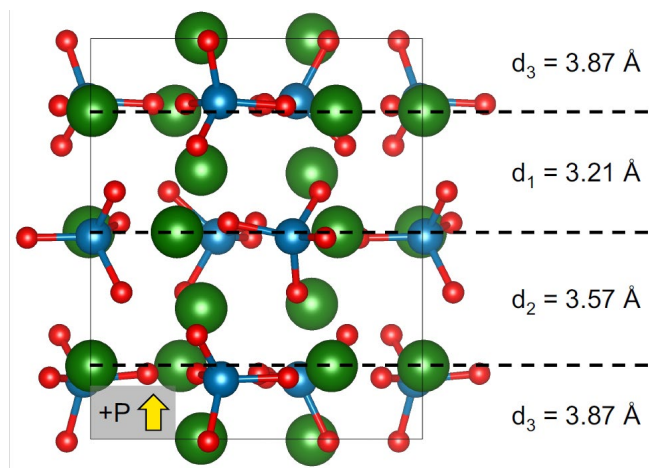

**Figure S8 | DFT calculated ferroelectric structure.** Calculated ferroelectric structure ($+P$) of $Pb_5Ge_3O_{11}$, and resulting Pb plane spacings $d_1$, $d_2$, and $d_3$.

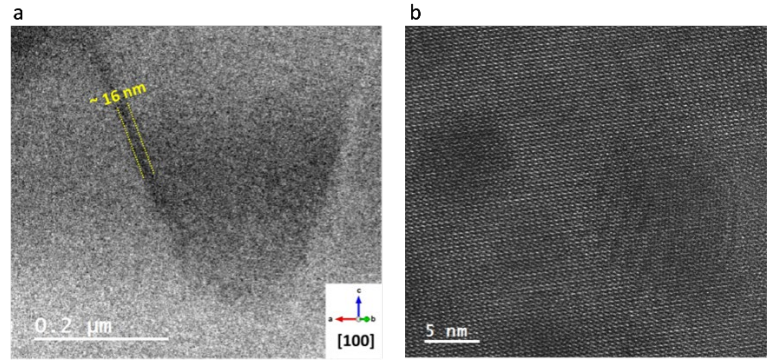

**Figure S9 | Bright field imaging of domains and domain walls. a,** Illumination at a relatively low dose ($\approx 33$ e-/Å$^{-2}$) and low magnification reduced effects from domain wall motion, revealing a region of reduced intensity in the transition region between $+P$ and $-P$ domains with a width of about 16 nm and larger (not shown). **b,** Imaging at higher magnification with the lowest dose possible in our microscope setup (see Methods for details) resulted in beam damage, prohibiting measurements with atomic resolution by bright field TEM.

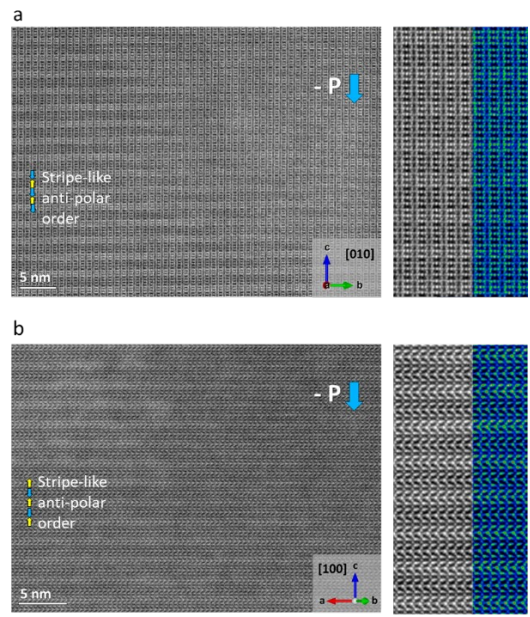

**Figure S10 | STEM imaging of a -*P* domain, transitioning into stripe-like anti-polar order towards a head-to-head domain wall. a,** STEM ADF imaging map of a -*P* domain transitioning into an aperiodic region, imaged along the [010] direction. A cropped region with higher magnification is present in the panel on the right, partly false coloured to highlight the emergent stripe-like pattern. **b,** Same as in **a**, imaged along the [100] direction in another sample.

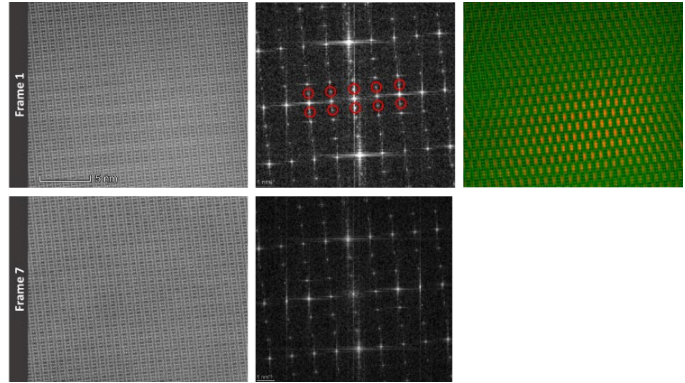

**Figure S11 | STEM FT masking of domain wall and domain regions.** Frame 1 corresponds to an STEM ADF frame taken with a domain wall in the field of view. The domain wall causes specific superlattice spots (circled in red), which vanish as the domain wall leaves the field of view. As show in the top right panel, FT masking of the domain-wall related spots highlights the stripe-like antiferroelectric region;red regions in the reverse FT are from the superlattice spots circled in red, detailed in Ref.[20]. By continuously scanning, the domain wall is moved out of the field of view and thus the superlattice reflections disappear (Frame 7).

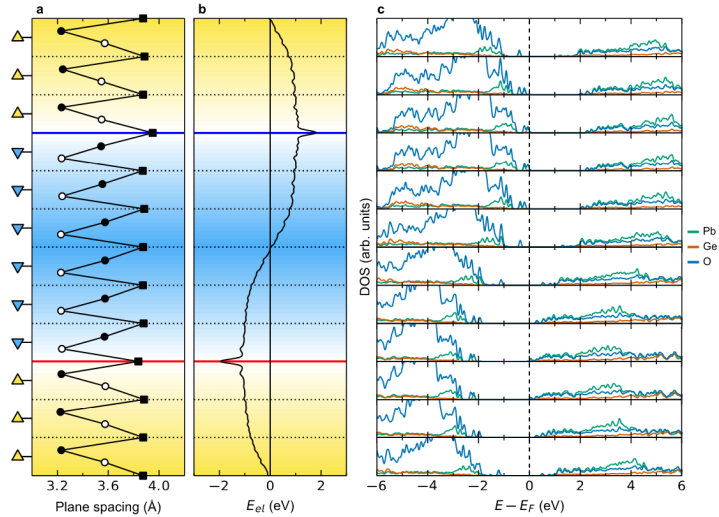

**Figure S12 | DFT modeling of conventional domain walls in $Pb_5Ge_3O_{11}$. a-c,** Calculated Pb plane spacing (**a**), local potential profile (**b**), and electronic density of states (DOS) (**c**) across the supercell. Yellow and blue arrow heads in **a** correspond to +*P* and -*P* polarizations, respectively. White circles, black circles, and squares in **a** correspond to the plane spacings $d_1$, $d_2$, and $d_3$, respectively. The horizontal lines in **a-b** illustrate positions of interfaces between unit cells of opposite polarization direction, where the red and blue lines mark the center of the charged domain walls. The vertical dashed line in **c** corresponds to the Fermi level.